# Naming the Identified Feature Implementation Blocks from Software Source Code

Ra'Fat Al-Msie'deen, Hamzeh Eyal Salman, Anas H. Blasi, and Mohammed A. Alsuwaiket

*Abstract*—Identifying software identifiers that implement a particular feature of a software product is known as feature identification. Feature identification is one of the most critical and popular processes performed by software engineers during software maintenance activity. However, a meaningful name must be assigned to the *Identified Feature Implementation Block* (IFIB) to complete the feature identification process. The feature naming process remains a challenging task, where the majority of existing approaches manually assign the name of the IFIB. In this paper, the approach called *FeatureClouds* was proposed, which can be exploited by software developers to name the IFIBs from software code. FeatureClouds approach incorporates word clouds visualization technique to name *Feature Blocks* (FBs) by using the most frequent words across these blocks. FeatureClouds had evaluated by assessing its added benefit to the current approaches in the literature, where limited tool support was supplied to software developers to distinguish feature names of the IFIBs. For validity, FeatureClouds had applied to draw shapes and ArgoUML software. The findings showed that the proposed approach achieved promising results according to well-known metrics in terms of Precision and Recall.

## I. INTRODUCTION

FEATURE identification or location is the process of detecting the source code elements, such as classes or methods, that implement particular functionality in a software product [1]. Several works have been carried out on feature identification, whether from single software or a group of software products [2]. Feature naming is the activity of suggesting a meaningful name for the extracted feature implementations. In this work, a feature is a functionality provided by a software product.

Software identifier name (*e.g.,* package, class, method, and attribute) is one of the most significant software understanding resources [3]. The identifier names of *Feature Implementation Block* (FIB) need to be analyzed for feature naming. Normally, a FIB contains many identifier names which include several words. Software identifier names are often constructed by mixing fragments of words, acronyms, and abbreviations (*e.g.* setRectangleY). For feature identification of a legacy software system, one main problem is to understand the FIB and name it.

The analysis of identifier names is a very helpful way in naming decisions. Current studies support the feature identification process in a single software [4], or in a set of software product variants [5]. Considering the feature identification process, the current approaches are based either on a static code analysis or on a dynamic analysis of the software product. Other approaches further exploit information retrieval methods [6]. In the use of the proposed approach, we suppose that the IFIBs exist in advance. Thus, it is important to clarify that the feature identification process is out of the scope of this study. This paper focuses only on the feature naming of the identified blocks. Figure 1 presents an example of the IFIBs from software source code.

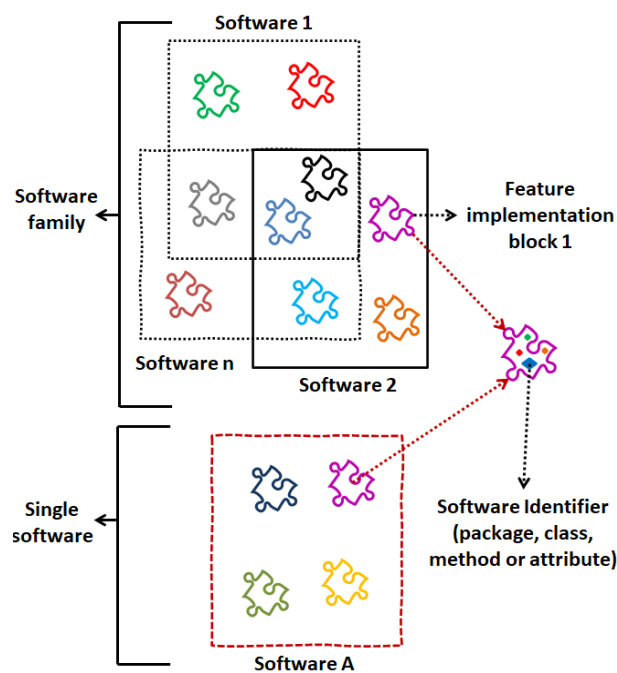

Fig. 1. The IFIBs from software source code

Figure 1 shows that the implementation of features can either be identified from a single software system or by a group of software product variants. Also, the given figure shows that each block that constitutes an implementation of the feature consists of a set of software identifiers such as package, class, method, and attribute.

Existing contributions in the literature of feature identification, including the work of Al-msie'deen *et al.* [5], overlook the feature naming phase during the identification of FIBs. The majority of current feature implementations identification approaches are not focused on a feature naming problem, and usually, this phase is carried out manually [7, 8]. The absence of support for feature naming in existing approaches is a critical threat to their efficiency.

In this paper, we propose an approach called *FeatureClouds* to support feature naming during the feature identification process from software source code. The proposed approach exploits the word cloud visualization technique [9] to help software engineers to name the IFIBs. As each block include only the implementation of a single feature. Feature naming is based on an analysis of the content of each IFIB to propose a feature name. Where the IFIBs can be named by selecting words with the highest frequency from word clouds.

FeatureClouds is a generic approach, where this approach works on the IFIBs whether from a single software product or software product variants. FIB is a software artifact the consists of a collection of software identifiers (*i.e.,* packages, classes, methods, or attributes). Thus, these code identifiers implement a particular software feature.

In our work, we do not suppose there is domain knowledge regarding the features of a software product. Therefore, the domain experts cannot manually suggest feature names to FIBs based on their knowledge about the existing software system. Thus, there is a need to develop an approach for feature naming without domain expert intervention and based on the vocabulary used in the identified blocks. FeatureClouds is a significant improvement over existing approaches, as it aims to automatically give a name for the IFIB based on its content.

The rest of the paper is structured as follows. Section II presents related work closest to the proposed contribution. Section III details FeatureClouds approach. Section IV presents FeatureClouds experimentations. Finally, Section V concludes the paper.

## II. RELATED WORK

Davril *et al.* [10] introduced a feature naming method as a part of their Feature Model (FM) extraction approach. The authors presented an approach for constructing a FM from product descriptions in natural language. In their work, to build the FM and present it to the human user, there is a need to assign a meaningful name for the identified feature (cluster or block). Thus, the authors developed a cluster-naming process to select the most frequent phrase from all feature descriptors in the cluster. Meaningful names are selected for the mined features based on the most frequent phrases discovered for each of the clusters or blocks.

Al-msie'deen *et al.* [11] used the source code of the mined FIBs to generate feature names automatically. In their work, all identifier names found in the FIB are mined. Then, each identifier name is divided into tokens by using a camel-case scheme. After that, a weight is assigned to each extracted token based on its frequency in the block. Finally, a feature name was constructed using the highest weighted words. The number of words used in the feature name was selected by a software engineer. For instance, the engineer can choose the highest two words to create the feature name. For the purpose of features readability, the authors assigned feature names based on the most frequent tokens of the IFIBs. They did not evaluate feature names and they did not provide details about the feature naming process.

Martinez *et al.* [12] offered a word cloud visualization technique to support software developers in naming the IFIBs from a collection of software product variants. This visualization is used through the feature implementation identification process to propose feature names to software engineers. In their work, once FIBs are identified from software product variants, the authors used the *VariClouds* approach to visualize the code elements inside each FIB and determine important words that assist software engineers to identify feature names. VariClouds approach employs information retrieval methods, like TF-IDF, to analyze the code elements inside each FIB. Their approach is semi-automatic, where the domain expert manually reviews words in the cloud to identify feature names for the IFIBs. While this study presents a fully automatic approach for feature naming based on the IFIBs in advance, without any interventions from the domain expert. We conducted experiments using a real FIBs to verify our intuition that word cloud gives better results in this field.

The study presented by Martinez *et al.* [12] is the closest to ours. The authors use optional filters such as *camel case splitter* for words dividing. In this paper, the word processing has done before it has presented in the final word cloud. This processing is done via word splitting and stemming. These two steps are the core of our approach and not an optional filter. FeatureClouds uses two filters to filter out unwanted words from the word clouds. The short-word filter intends to filter out the words which have less than three letters. Also, the word-frequency filter is used as a sign for the word frequency across the IFIBs. In the VariClouds approach, there is no indication of how many times a word is repeated within a block. Word cloud layout in VariClouds approach is a typewriter. While FeatureClouds layouts are typewriter and spiral layouts [9]. VariClouds displayed words in the word cloud in alphabetical order (*i.e.,* a-z). While FeatureClouds shows words in the cloud in an alphabetical or frequency order. In frequency order, words appear according to their importance in the cloud where the most important words appear first in the word cloud. VariClouds did not provide clear evaluation criteria for the quality of the feature name obtained for each block. While we evaluated our results with clear metrics like recall and precision.

AL-msie'deen *et al.* [7] assigned manually the feature names to the IFIBs, based on the study and analysis of the content of each block and on their knowledge of software product variants. Where the software variants are well documented, and their feature names are known in advance. AL-msie'deen *et al.* [8] suggested feature naming process as a research direction to suggest the feature names automatically for the IFIBs. Consequently, that is what we have done in this study by developing a feature naming approach called FeatureClouds.

*Ziadi et al.* [13] suggested an approach to find feature implementations across software variants source code. The authors manually proposed feature names of the IFIBs. The names were given to feature implementations based on the existing FM document. Thus, our work is more effective for feature naming, where we automatically assign feature names of the IFIBs based on the word cloud.

AL-msie'deen *et al.* [14] developed an approach to suggest a name for the IFIBs based on the use-case diagrams of software variants. In their study, FIB has given a name based on the textual similarity between the use-case description and block content. In our work, we rely just on the content of the FIB, and we do not need any other artifacts of the software system. Furthermore, our work considers the legacy software system, which is not documented well, and usually, all software artifacts are missing.

*Adjoyan et al.* [15] suggested an automatic approach for *service* identification from the Object-Oriented (OO) source code. Their paper aims to migrate the OO legacy system into Service Oriented Architecture (SOA). For the legibility of the identified service blocks, authors assign names based on the most frequent words across the identified blocks. The authors documented the resulting services by allocating a name using the most frequent words in their class names. The authors do not evaluate service names, and they do not offer details about the service naming process. The proposed approach can be used to name the identified service implementation blocks.

*Kebir et al.* [16] suggested a method to extract *components* from software source code. The authors suggested names to the identified components based on the source code of the identified clusters. They provided component names based on the most frequent tokens of the identified clusters. For each cluster, the class names are split into words based on the camel-case method. Then, a weight is given to each obtained token, and at last, a component name is created using the strongest weighted tokens. FeatureClouds approach can be used to name the identified component implementation clusters.

*Shatnawi et al.* [17] proposed an approach to reverse engineer the architecture model of a collection of software product variants. They aimed to identify the main components and dependencies between those components. In their work, for comprehensibility, they named the identified components by using the most frequent words across the identified component implementation clusters. Each component cluster contains a collection of software classes. Based on the most frequent words across class names, they allocate a name for that cluster. Their work is very similar to the study proposed by Adjoyan *et al.* [15]. Also, the authors of this study do not evaluate component names, and they do not give any details about the component naming process. The approach proposed in this paper can be applied to name the identified component implementation clusters.

Table I presents a comparison between feature naming approaches. The studied approaches have been evaluated based on the criteria of naming method (automatic versus manual), inputs (feature blocks, product descriptions, service blocks, component blocks, use-case diagrams), and outputs (word clouds, most frequent tokens, feature names).

TABLE I
SUMMARY OF FEATURE NAMING APPROACHES (COMPARISON TABLE)

| ID | Reference | Input | | | | | Output | | | Naming method | | |
|---|---|---|---|---|---|---|---|---|---|---|---|---|
| | | Feature blocks | Product descriptions | Service blocks | Component blocks | Use-case diagrams | Word clouds | Most frequent tokens | Feature names | Manual | Semi-automatic | Automatic |
| 1 | Davril *et al.* [10] | | x | | | | | x | | | | x |
| 2 | Al-msie'deen *et al.* [11] | x | | | | | | x | | | | x |
| 3 | Martinez *et al.* [12] | x | | | | | x | | | | x | |
| 4 | AL-msie'deen *et al.* [7] | x | | | | | | x | x | | | |
| 5 | Ziadi *et al.* [13] | x | | | | | | x | x | | | |
| 6 | AL-msie'deen *et al.* [14] | x | | | | x | | x | | | | x |
| 7 | Adjoyan *et al.* [15] | | | x | | | | x | | | | x |
| 8 | Kebir *et al.* [16] | | | | x | | | x | | | | x |
| 9 | Shatnawi *et al.* [17] | | | | x | | | x | | | | x |
| 10 | Al-msie'deen *et al.* [FeatureClouds] | x | | | | | x | | | | | x |

The majority of the current studies manually suggest feature names of the extracted FIBs based on existing software documentation. In the related work, no work automatically provides a name for the IFIBs.

Table II presents a comparison between feature naming approaches that exploited the word cloud visualization technique. There is only one study concerned with feature naming based on the word-cloud in the literature, which is the study of Martinez *et al.* [12]. We evaluate this closest work to our approach based on the following criteria: granularity level of block code, programmed method, cloud filters, evaluation criteria of the proposed name, cloud layout, cloud arrangement, word preprocessing.

TABLE II
SUMMARY OF FEATURE NAMING APPROACHES BASED ON WORD CLOUD TECHNIQUE (COMPARISON TABLE)

| Reference | Feature blocks - code granularity level | | | | Programmed method | | Cloud filters | | | | Evaluation criteria | | Cloud layout | | Cloud arrangement | | Word preprocessing | |
|---|---|---|---|---|---|---|---|---|---|---|---|---|---|---|---|---|---|---|
| | Package | Class | Method | Attribute | Semi-automatic | Automatic | Camel-case splitter | Words stemming | Short word | Word-frequency | Recall | Precision | Typewriter | spiral | Alphabetical | Frequency | Words splitting | Words stemming |
| Martinez [12] | | x | | | x | | x | x | | | | | x | | x | | | |
| FeatureClouds | x | x | x | x | | x | x | x | x | x | x | x | x | x | x | x | x | x |

## III. THE FEATURECLOUDS APPROACH

An overview of the suggested method is given in Figure 2. The inputs are the FIBs extracted from software source code. The outputs are the most frequent words across FIBs (*i.e.,* FIB names).

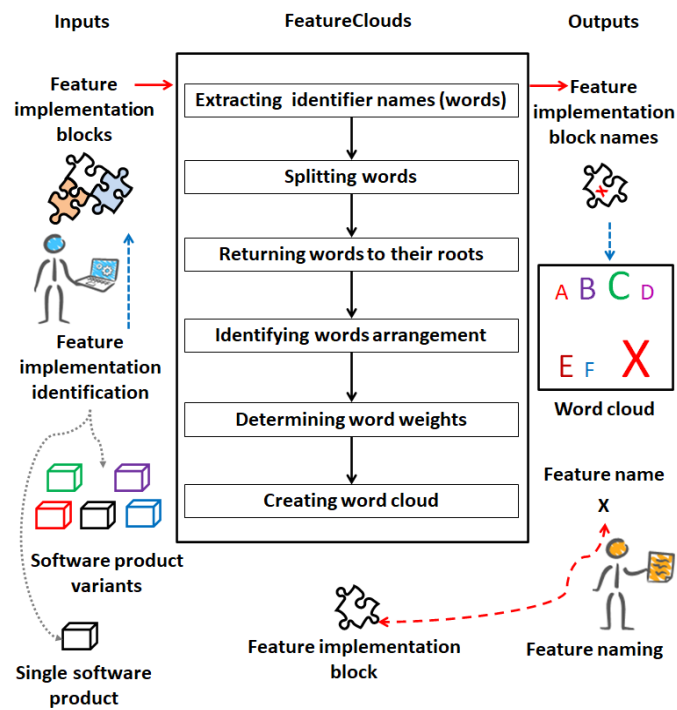

Fig. 2. Overview of FeatureClouds process

FIB contains the source code elements (*aka.* block identifiers) that implement a particular feature. Thus, from each block, FeatureClouds extracts the block identifier names. The block identifier names are the only source of the naming process. Figure 3 shows an example of FIBs identified from *drawing shapes* software variants by the *Revpline* approach [18]. The IFIBs consist of all code granularity levels (*i.e.,* package, class, method, and attribute).

In this paper, we rely on the block contents to assign a feature name for each block. Feature implementations are blocks of the code identifiers. Word clouds are a representation of the identifier names that are constructed in the FIBs. These word clouds are built with a FeatureClouds approach as detailed in the following.

### A. Extracting Identifier Names (words)

The first step of our feature naming process is the extraction of software identifier names (*aka.* words) from the IFIB. Table III shows the software identifier names for each FB in Figure 3.

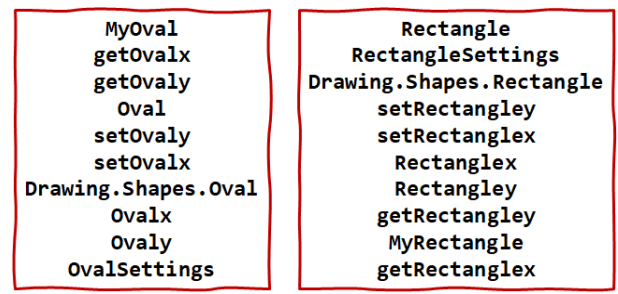

Fig. 3. FIBs identified from drawing shapes software variants

FeatureClouds accepts the IFIB as input. Then, FeatureClouds produces a words file as output for each block. The words file of a particular block includes all software identifier names of this block.

TABLE III
IDENTIFIER NAMES EXTRACTED FROM FIBS IN FIGURE 3

| Identifier names | Identifier names |
|---|---|
| MyOval | Rectangle |
| getOvalx | RectangleSettings |
| getOvaly | Drawing.Shapes.Rectangle |
| Oval | setRectangley |
| setOvaly | setRectanglex |
| setOvalx | Rectanglex |
| Drawing.Shapes.Oval | Rectangley |
| Ovalx | getRectangley |
| Ovaly | MyRectangle |
| OvalSettings | getRectanglex |

### B. Splitting Words

In this step, software identifier names are split into words (or tokens) based on the camel-case syntax [18]. For example, *RectangleSettings* is split into *rectangle* and *settings*. Camel-case method splits identifier names based on capital letters, special characters, and numbers. This method is uncomplicated and commonly used for software identifier splitting. Table IV shows samples of the split words from the drawing shapes software identifiers.

TABLE IV
WORDS OBTAINED FROM FIBS IN FIGURE 3

| Words or tokens | Words or tokens |
|---|---|
| my | rectangle |
| oval | rectangle |
| get | settings |
| ovalx | drawing |
| get | shapes |
| ovaly | rectangle |
| oval | set |
| set | rectangley |
| ovaly | set |
| set | rectanglex |
| ovalx | rectanglex |
| drawing | rectangley |
| shapes | get |
| oval | rectangley |
| ovalx | my |
| ovaly | rectangle |
| oval | get |
| settings | rectanglex |

### C. Returning Words to Their Roots

In this step, the word stemming process is performed (*e.g.,* eliminating word endings) via WordNet [19]. WordNet tool is a huge lexical database of the English language. In our work, stemming is employed to replace English words with their stems or roots. For instance, in drawing shapes software, the root of the word "drawing" is "draw". Table V shows samples of the word roots from drawing shapes software.

TABLE V
SAMPLES OF ENGLISH WORDS AND THEIR ROOTS FROM DRAWING SHAPES

| Word roots | Word roots |
|---|---|
| My | Rectangle |
| Oval | Rectangle |
| Get | Set |
| Ovalx | Draw |
| Get | Shape |
| Ovaly | Rectangle |
| Oval | Set |
| Set | Rectangley |
| Ovaly | Set |
| Set | Rectanglex |
| Ovalx | Rectanglex |
| Draw | Rectangley |
| Shape | Get |
| Oval | Rectangley |
| Ovalx | My |
| Ovaly | Rectangle |
| Oval | Get |
| Set | Rectanglex |

### D. Identifying Words Arrangement

FeatureClouds employs *typewriter*-style to place words in the word cloud from left to right side, and from top to bottom side. FeatureClouds shows words in the word cloud in alphabetical arrangement. Software engineer appears more capable to find words in alphabetically ordered word clouds [9]. Table VI displays samples of words in an alphabetical arrangement.

### E. Determining Word Weights

In this step, the weight is given to the word based on its frequencies in the IFIB. In our work, the weight of the word shows the word frequency in a given FB. Table VII shows samples of words and their weights from drawing shapes software.

TABLE VI
SAMPLES OF WORDS IN AN ALPHABETICAL ARRANGEMENT

| Words in alphabetical order | Words in alphabetical order |
|---|---|
| Draw | Draw |
| Get | Get |
| Get | Get |
| My | My |
| Oval | Rectangle |
| Oval | Rectangle |
| Oval | Rectangle |
| Oval | Rectangle |
| Ovalx | Rectanglex |
| Ovalx | Rectanglex |
| Ovalx | Rectanglex |
| Ovaly | Rectangley |
| Ovaly | Rectangley |
| Ovaly | Rectangley |
| Set | Set |
| Set | Set |
| Set | Set |
| Shape | Shape |

TABLE VII
SAMPLES OF WORDS AND THEIR WEIGHTS FROM DRAWING SHAPES SOFTWARE

| Word | Weight | Word | Weight |
|---|---|---|---|
| Draw | 1 | Draw | 1 |
| Get | 2 | Get | 2 |
| My | 1 | My | 1 |
| Oval | 4 | Rectangle | 4 |
| Ovalx | 3 | Rectanglex | 3 |
| Ovaly | 3 | Rectangley | 3 |
| Set | 3 | Set | 3 |
| Shape | 1 | Shape | 1 |

Word weight is a very important issue in our approach, as repeating a given word multiple times in a FIB is a good indication of the importance of that word. When a software developer uses one word to name several software identifiers, this indicates the importance of this word. Mostly, a word that is repeated frequently within a given block reflects the functionality that that block provides to the end-user. For example, in the drawing shapes software, when the word "*oval*" is repeated more than the rest of the words in the block (4 times), this block is named *oval*. The function provided by this FB is to enable the drawing of an oval by the user of this software. As a result, there is a close relationship between the functionality provided by the given block and the most frequently occurring words across it. Therefore, the word with a higher weight is suitable for naming the given block.

### F. Creating Word Cloud

FIBs are obtained using existing feature identification approaches such as the *Revpline* approach [18]. Then, the software engineers use *FeatureClouds* to assign a feature name to each block. The name is assigned based on the most frequent identifier name in each block, where the constructed word clouds show the most frequent words across each block. The main *hypothesis* of our approach for feature naming is that the most frequent words across each block are those that make each feature implementation unique regarding the rest of the IFIBs. Also, in our work, we consider that the most frequent words across each block reflect the real functionality provided by each block to the end-user (*i.e.,* feature). Figure 4 shows the word clouds that are constructed from the FIBs in Figure 3. In Figure 4, the *word-frequency* filter can be used as a pointer for the word frequency across FIB. This filter defines the frequency of the word across the block as an accurate number between square brackets after any word in the cloud.

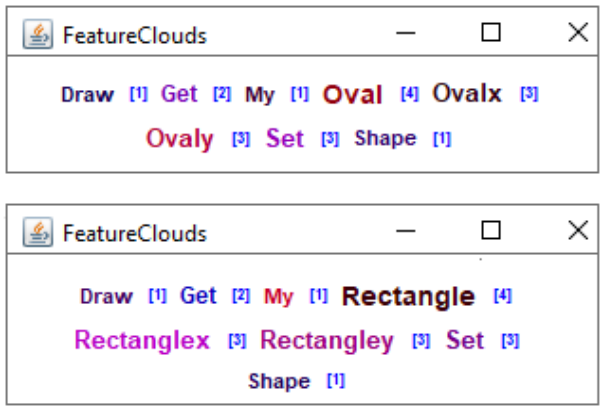

Fig. 4. Word clouds built from the IFIBs in Figure 3

In Figure 4, FeatureClouds assigns the caption (*aka.* label) "*oval*" to the IFIB. Also, it assigns the caption "*rectangle*" to the other FIB. The number of words used in a feature name is selected by the software developer, where he can change the settings of the FeatureClouds approach to retrieve the highest two (or three) words for each block.

In the drawing shapes software case study, the name of the features is well documented through the FM [20]. The FM gives us a ground truth to assess the feature naming process proposed in this work, where the manual feature naming is presented in the drawing shapes software FM [18]. The real name of the oval feature in the FM is "*draw_oval*", while the name of the rectangle feature is "*draw_rectangle*". The font size of the word in the retrieved word cloud is the number of times the word is repeated throughout the IFIB. In the word clouds, words that emerge with a large font size are more critical than the rest of the words.

## IV. EXPERIMENTATION

This section presents the ArgoUML case study, evaluation metrics, experimental results, and the threats to validity of our approach.

### A. Case Study

To evaluate the FeatureClouds approach, we selected the ArgoUML case study, where the name of the features are well known and documented. ArgoUML is an open-source, Java-based program. ArgoUML variants are ten software products, and its FM consists of nine features [21]. These features are: *class*, *activity*, *collaboration*, *use-case*, *state*, *sequence*, *cognitive*, *logging*, and *deployment* feature. We recommend researchers use the *Revpline* approach to extract FIBs from a collection of software product variants [18]. Figure 5 shows the ArgoUML FM.

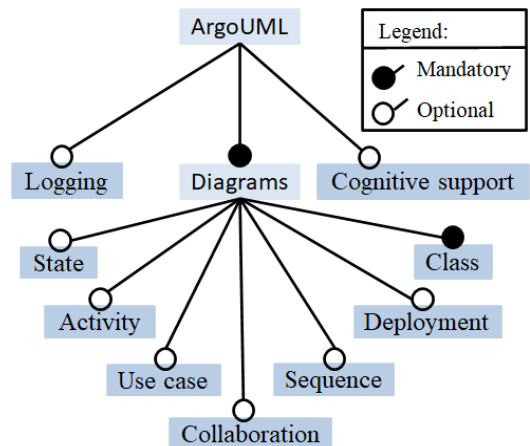

Fig. 5. ArgoUML FM [21]

### B. Evaluation Metrics

The effectiveness of our approach is determined by their *recall*, and *precision* [18]. For the named feature by FeatureClouds, a precision metric is the proportion of correctly retrieved words from the word cloud to the total number of retrieved words from the word cloud (*cf.* Equation 1). Recall metric is the proportion of correctly retrieved words from word cloud to the total number of relevant words from the manual feature name (*cf.* Equation 2). All FeatureClouds metrics have values between zero and one. If the recall is equal to one, all relevant feature name words are retrieved. However, some retrieved words might not be relevant to the manual feature name. If precision is equal to one, all retrieved feature name words are relevant. However, relevant words might not be retrieved from a word cloud. The *evaluation metrics* of FeatureClouds approach for a feature name are defined as follows:

$$\text{Precision} = \frac{\sum \text{correctly retrieved words}}{\sum \text{words that are retrieved}} \quad (1)$$

$$\text{Recall} = \frac{\sum \text{correctly retrieved words}}{\sum \text{words that are relevant}} \quad (2)$$

### C. Experimental Results

Figure 6 shows the IFIB, at class level, of activity diagram feature from ArgoUML variants. The real name of this feature is "*activity*". FeatureClouds retrieves two words for this block as feature name, which are *activity* and *diagram*. Our implementation of *FeatureClouds* is available at the main author website [22].

Figure 7 shows the word cloud extracted from the activity diagram implementation block. The most frequent word across this block is activity and diagram. Software engineers might not use meaningful vocabularies to name the software identifiers. In this case, the FeatureClouds approach will fail to provide a meaningful name for the identified FB. For example, the retrieved feature name "*class diagram*" has more meaning than the "*action model*" name.

```
                    FigPool
         ActivityDiagramGraphModel
                FigCallState
             FigSubactivityState
          ActivityDiagramLayouter
              SelectionCallState
              SelectionPartition
             FigObjectFlowState
                FigPartition
               FigActionState
      PropPanelUMLActivityDiagram
            InitActivityDiagram
          ActionActivityDiagram
            ActionCreatePartition
          ActivityDiagramRenderer
      ActivityDiagramPropPanelFactory
              ModePlacePartition
             UMLActivityDiagram
```

Fig. 6. The IFIB of the activity diagram feature

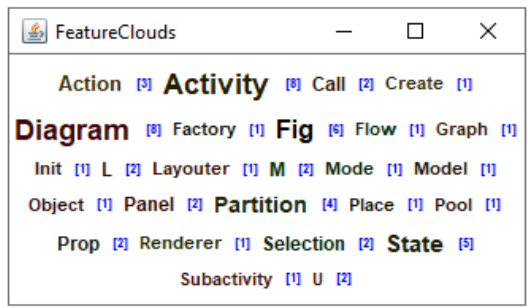

Fig. 7. A word cloud of the IFIB In Figure 6

FeatureClouds gives a "*collaboration*" name for the IFIB of the collaboration feature in ArgoUML. Also, FeatureClouds assigns a "*state*" name for the IFIB of the state feature from ArgoUML. In our work, when naming all FIBs, it is impossible that two or more FIBs have the same name. All the named features from ArgoUML have a *unique* name.

Table VIII shows the feature names named by the FeatureClouds approach and shows the manual names of ArgoUML features as described in the ArgoUML FM. Results show that the FeatureClouds approach can retrieve a feature name for the IFIB. The feature name suggested by the approach is too close to the real or manual name, as described in the FM.

Moreover, the experimental results in Table VIII show that the word clouds are useful for assigning the names for FIBs. Especially when manual feature names were not available, word clouds consider a unique method and perfect technique toward feature naming. Results also proved that the word clouds worked as confirmation for the feature naming decision. Based on the obtained results, we can say that word clouds minimize software engineers' understanding time and assist them to be more reliable with the feature naming decisions.

Depending on the results obtained from the presented case study, we should state here that the words that formed the manual name of the *class* and *cognitive support* features are not retrieved by the suggested approach as the feature name. On the other hand, these words appear undoubtedly on the word cloud (*e.g.,* class and cognitive). These words appear less frequently than the other words in the block. Thus, the approach doesn't assign it as the feature name of the IFIBs.

TABLE VIII
FEATURE NAMING RESULTS IN ARGOUML

| Feature Naming Findings in ArgoUML Variants | | Evaluation Metrics | |
|---|---|---|---|
| Feature naming via the FM | Feature naming via FeatureClouds approach | Recall | Precision |
| State | State | 100% | 100% |
| Collaboration | Collaboration | 100% | 100% |
| Activity | Activity & diagram | 100% | 50% |
| Use case | Use & case | 100% | 100% |
| Sequence | Fig & sequence & diagram & message | 100% | 25% |
| Deployment | Fig & deployment & diagram | 100% | 33% |
| Class | Action & m & u & l & model & list | 0% | 0% |
| Cognitive support | Cr & to & name & do | 0% | 0% |
| Logging | Log & info | 100% | 50% |

In ArgoUML, there is a mismatch between the manual names and the implementation details of some features. For example, in the case of the *cognitive support* feature, there is a full mismatch between the manual name and the words (or vocabulary) arising from FIB. The most frequent words of the cognitive support block are: *cr*, *to*, *name*, and *do* (*cf.* Table VIII). Moreover, there is a full mismatch between the manual name and the words appearing from FIB of a *class* diagram. The most frequent words of the class block are: *action*, *m*, *u*, *l*, *model*, and *list* (*cf.* Table VIII).

In the ArgoUML case study, findings show that *recall* appears very high for the majority of retrieved feature names by FeatureClouds (*cf.* Table VIII). This means that all words formed the manual feature name are retrieved via word cloud. For the class and cognitive support features, the recall metric is equal to zero. This means that the approach was unable to retrieve the feature name (or the words that make up the feature name) from the word cloud for these two features. Considering the *precision* metric, it is also quite high thanks to our FeatureClouds approach that identifies feature names based on the vocabulary of the IFIBs. Figure 8 shows the precision and recall for *state* feature. For cognitive support and class features, the precision metric is equal to zero. This means that our approach was unable to retrieve the feature names for these blocks, where the relevant words are not retrieved from the word cloud.

For qualitative analysis of our approach, we evaluated FeatureClouds with three software engineers familiar with ArgoUML. Software engineers performed a feature naming process for the IFIB by using word clouds. We asked to report their feedback for FB naming. Engineers found that feature naming using the word clouds consider an excellent method where the retrieved cloud represents all words and their frequencies for each block. Also, engineers found that the use of the word cloud visualization technique has proven helpful in supporting domain experts with feature naming, especially when domain knowledge is missing. In addition, engineers stated that the word cloud visualization paradigm is an effective technique for feature naming and very helpful in the naming decisions and accelerating the feature naming process.

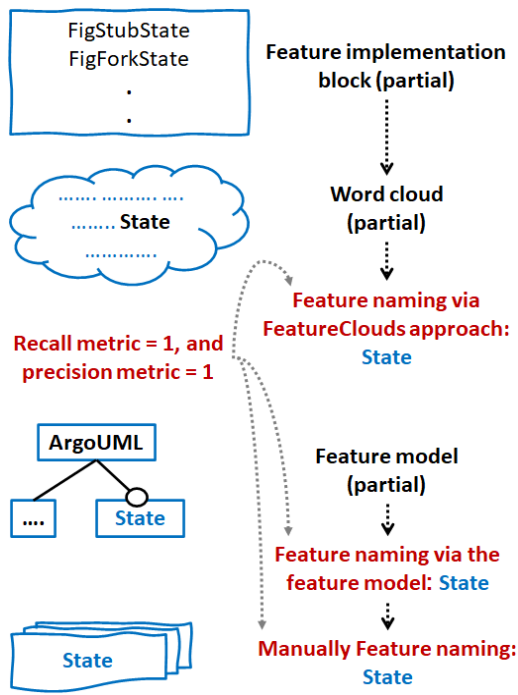

Fig. 8. FIB, word cloud, manual feature name, named feature by FeatureClouds, recall and precision metrics for state feature of ArgoUML software

The results found that the best straightforward method for extracting the feature name from a block of identifier names is to split identifier names into fundamental words (*e.g.,* OvalSettings → Oval + Settings). Also, to make the feature name more clear, fundamental words must be returned to their roots or stems (*e.g.,* Logging → Log). In our work, when a software identifier name is developed of partial mixed words, the camel-case splitting algorithm is no longer useful and should be enhanced with other methods. For instance, gET_OptimizeN is split into "*g*", "*E*", "*T*", "*Optimize*", and "*N*".

To evaluate the performance of our approach, we selected a large software system as a case study where the names of the software features are well documented. FeatureClouds shows excellent performance during the feature name process based on the execution times. Also, the results show the scalability of our approach to dealing with large, medium, and small systems. The ArgoUML case study is considered a large software system, where it is consists of 120,348 Lines of Code (LoC). Table IX presents feature naming results of the ArgoUML case study in more detail. The findings are characterized by metrics NoC (Number of Classes), NoW (Number of Words), ET (Execution Times) in ms, and MFW/R (Most Frequent Words/ Repetition).

TABLE IX
FEATURE NAMES DETAILS FOR ARGOUML SOFTWARE BASED ON FEATURECLOUDS

| ID | Feature name | NoC | NoW | ET | MFW/R | |
|----|--------------|-----|-----|------|-------|---|
| 1 | State | 35 | 50 | 1303 | State (32) | Fig (15) |
| 2 | Collaboration | 16 | 25 | 963 | Collaboration (12) | Diagram (9) |
| 3 | Activity | 18 | 26 | 980 | Activity (8) | Diagram (8) |
| 4 | Use case | 39 | 33 | 1019 | Use (17) | Case (17) |
| 5 | Sequence | 38 | 47 | 1091 | Fig (16) | Sequence (12) |
| 6 | Deployment | 20 | 23 | 990 | Fig (8) | Deployment (7) |
| 7 | Class | 1494 | 581 | 10340 | Action (317) | Model (260) |
| 8 | Cognitive support | 205 | 198 | 3279 | Cr (86) | To (35) |
| 9 | Logging | 0 | 6 | 668 | Log (2) | Info (2) |

FeatureClouds can be used to name the identified feature, component, and service implementation blocks. Thus, FeatureClouds is a general approach and applicable to naming software features, components, and services. Also, FeatureClouds can be used to name the concepts extracted from the artifacts of the software system [23]. Moreover, FeatureClouds can be used to get the important vocabulary of the obtained evolution scenarios of the software system [24].

Comparing our approach to the work of Martinez *et al.* [12], which is the only work in the literature that addresses the feature naming process for the IFIBs based on the word cloud, we found the performance of our approach is better than their work. The proposed approach deals with all code granularity levels, while their approach deals with software classes only. The work of Martinez *et al.* [12] is a semi-automatic approach, where the engineer manually analysis the words of the cloud to identify feature names for the IFIB, while our work is an automatic approach, where the suggested approach automatically retrieves feature name to the IFIB without the intervention of domain experts.

Also, the work of Martinez *et al.* [12] doesn't show the repetition for each word in the given block. While our work shows the repetition for each word across a given block. Moreover, our approach includes preprocessing of software identifiers such as word splitting and stemming, while Martinez *et al.* [12] deal with software identifiers as it without any preparation process. Furthermore, the mined word clouds in the work of Martinez *et al.* [12] are missing cloud filters, while our clouds include unique filters such as short word and word-frequency filters.

*D. Threats to Validity*

The *threat to validity* of FeatureClouds approach is that software developers might not use a good vocabulary to name software identifiers (*i.e.,* identifiers are not properly named). This means that word cloud may not be trustworthy in all cases to assign a meaningful name to the IFIBs. Also, naming the feature using the identifier names of the IFIB is not always dependable. In the FeatureClouds approach, we rely on the most frequent words in the word cloud to suggest the name for each FIB. The proposed name may not be appropriate to the feature role or functionality. This means that identifier names are not suitable in all cases to retrieve feature name and should be enhanced with other techniques. Moreover, when a software engineer makes use of mixed words to label software identifiers (such as MyRecTanGle) the camel-case splitting method can't handle these identifiers and should be improved with other splitting algorithms. Furthermore, WordNet may not be trustworthy in all cases to return the word root and should be enhanced with other methods. Finally, a word cloud is missing important filters such as *search* filter and the cloud should be enhanced with other filters.

## V. CONCLUSION AND PERSPECTIVES

FeatureClouds is an approach that employs a word cloud visualization technique to provide feature names for the IFIBs from a set of product variants or single software. It is constructed for assisting software engineers in feature naming. We evaluated it in numerous case studies such as ArgoUML and drawing shapes software. The findings show its soundness in feature naming. The findings of FeatureClouds have shown some limitations for the feature naming process. For instance, the suggested approach has returned irrelevant names to some blocks, and this occurs when identifiers are not properly named by software programmers. Thus, the retrieved feature name may not be appropriate or reflect feature role or functionality in the software system. Also, WordNet or camel-case method may not be trustworthy in all cases to return the word root or to split identifier name. Thus, these methods should be enhanced and improved with other methods. In the current approach, we give the same weights for all software identifiers types that make up the IFIB. As future work direction, we plan to assess the use of word weights for different software identifier types (*i.e.,* package, class, method, and attribute). For instance, in the IFIB, the word that belongs to the class name has more importance (*i.e.,* weight) than the word that belongs to the attribute name.

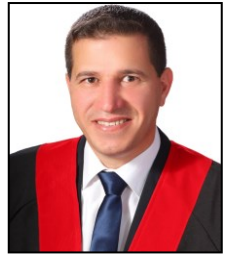

**Ra'Fat Al-Msie'Deen** is an Associate Professor in the Software Engineering department at Mutah University since 2014. He received his PhD in Software Engineering from the Université de Montpellier, Montpellier – France, in 2014. He received his MSc in Information Technology from the University Utara Malaysia, Kedah – Malaysia, in 2009. He got his BSc in Computer Science from Al-Hussein Bin Talal University, Ma'an – Jordan, in 2007. His research interests include software engineering, requirements engineering, software product line engineering, feature Identification, word cloud, and formal concept analysis.

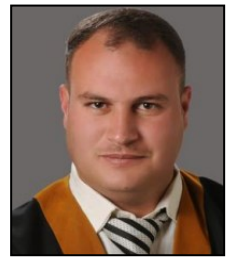

**Hamzeh Eyal Salman** is an Associate Professor at Mutah University since 2014. He received his PhD in Software Engineering from the Université de Montpellier, Montpellier – France, in 2014. He received his MSc in Computer Science from the University of Jordan, Amman– Jordan, in 2010. He got his BSc in Computer Information Systems from Al-Hussein Bin Talal University, Ma'an – Jordan, in 2006. His research interests include software engineering, software product line engineering, and formal concept analysis. He has published several papers in reputed journals and conferences.

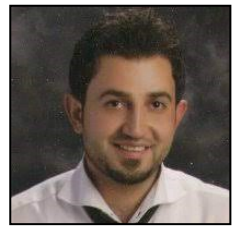

**Anas H. Blasi** is an Associate professor in the Data Science department at Mutah University. He earned the MSc in Computer Science from University of Sunderland (UK) in 2010, and the Ph.D. in Computer and systems Science from the State University of New York at Binghamton (USA) in 2013. Dr. Blasi research area is focusing on AI, Data Mining, Data Science, Machine Learning, Optimization algorithms, Fuzzy logic, and EDM. He has published several papers in reputed journals and conferences.

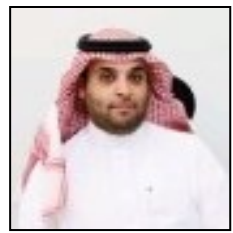

**Mohammed A. Alsuwaiket** is an Assistant professor in the Computer Science and Engineering Technology department at University of Hafr Al-Batin. He earned the MSc in Computer Science from University of Hertfordshire (UK) in 2012, and the Ph.D. in Computer Science from Loughborough University (UK) in 2018. Dr. Alsuwaiket research area is focusing on AI, Data Mining, Machine Learning, Fuzzy logic, and EDM. He has published several papers in reputed journals and conferences.